\documentclass[aps,twocolumn,preprintnumbers,amsmath,amssymb]{revtex4}
\usepackage{graphicx}
\usepackage{longtable}
\usepackage{dcolumn}
\usepackage{bm}
\usepackage{color}
\usepackage{sidecap}
\sidecaptionvpos{figure}{t}

\usepackage{hyperref}

\begin{document}
	
\title{Half-Metal Spin-Gapless Semiconductor Junctions as a Route to the Ideal Diode}

\author{Ersoy \c{S}a\c{s}{\i}o\u{g}lu$^{1}$}
\author{Thorsten Aull$^{1}$}
\author{Dorothea Kutschabsky$^{1}$}
\author{Stefan Bl\"{u}gel$^{2}$}
\author{Ingrid Mertig$^{1,3}$}

\affiliation{$^{1}$Institute of Physics, Martin Luther University Halle-Wittenberg, D-06120 Halle (Saale), Germany  \\
 $^{2}$Peter Gr\"{u}nberg Institut and Institute for Advanced Simulation, Forschungszentrum J\"{u}lich and JARA, D-52425 J\"{u}lich, Germany  \\
  $^{3}$Max Planck Institute of Microstructure Physics, Weinberg 2, D-06120 Halle (Saale), Germany}

\begin{abstract}
The ideal diode is a theoretical concept that completely conducts the electric current under forward bias without any loss and that behaves like a perfect insulator under reverse bias. However, real diodes have a junction barrier that electrons have to overcome and thus they have a threshold voltage $V_T$, which must be supplied to the diode to turn it on. This threshold voltage gives rise to power dissipation in the form of heat and hence is an undesirable feature. In this work, based on half-metallic magnets (HMMs) and spin-gapless semiconductors (SGSs) we propose a diode concept that does not have a junction barrier and the operation principle of which relies on the spin-dependent transport properties of the HMM and SGS materials. We show that the HMM and SGS materials form an Ohmic contact under any finite forward bias, while for a reverse bias the current is blocked due to spin-dependent filtering of the electrons. Thus, the HMM-SGS junctions act as a diode with zero threshold voltage $V_T$, and linear current-voltage (\textit{I-V}) characteristics as well as an infinite on:off ratio at zero temperature. However, at finite temperatures, non-spin-flip thermally excited high-energy electrons as well as low-energy spin-flip excitations can give rise to a leakage current and thus reduce the on:off ratio under a reverse bias. Furthermore, a zero threshold voltage allows one to detect extremely weak signals and due to the Ohmic HMM-SGS contact, the proposed diode has a much higher current drive capability and low resistance, which is advantageous compared to conventional semiconductor diodes. We employ the nonequilibrium Green’s function method combined with density-functional theory to demonstrate the linear \textit{I-V} characteristics of the proposed diode based on two-dimensional half-metallic Fe/MoS$_2$ and spin-gapless semiconducting
VS$_2$ planar heterojunctions.
\end{abstract}
\maketitle

\section{Introduction \label{sec:Intro}}

\begin{figure}[!b]
\begin{center}
\vspace{-0.76cm}
\includegraphics[scale=0.36]{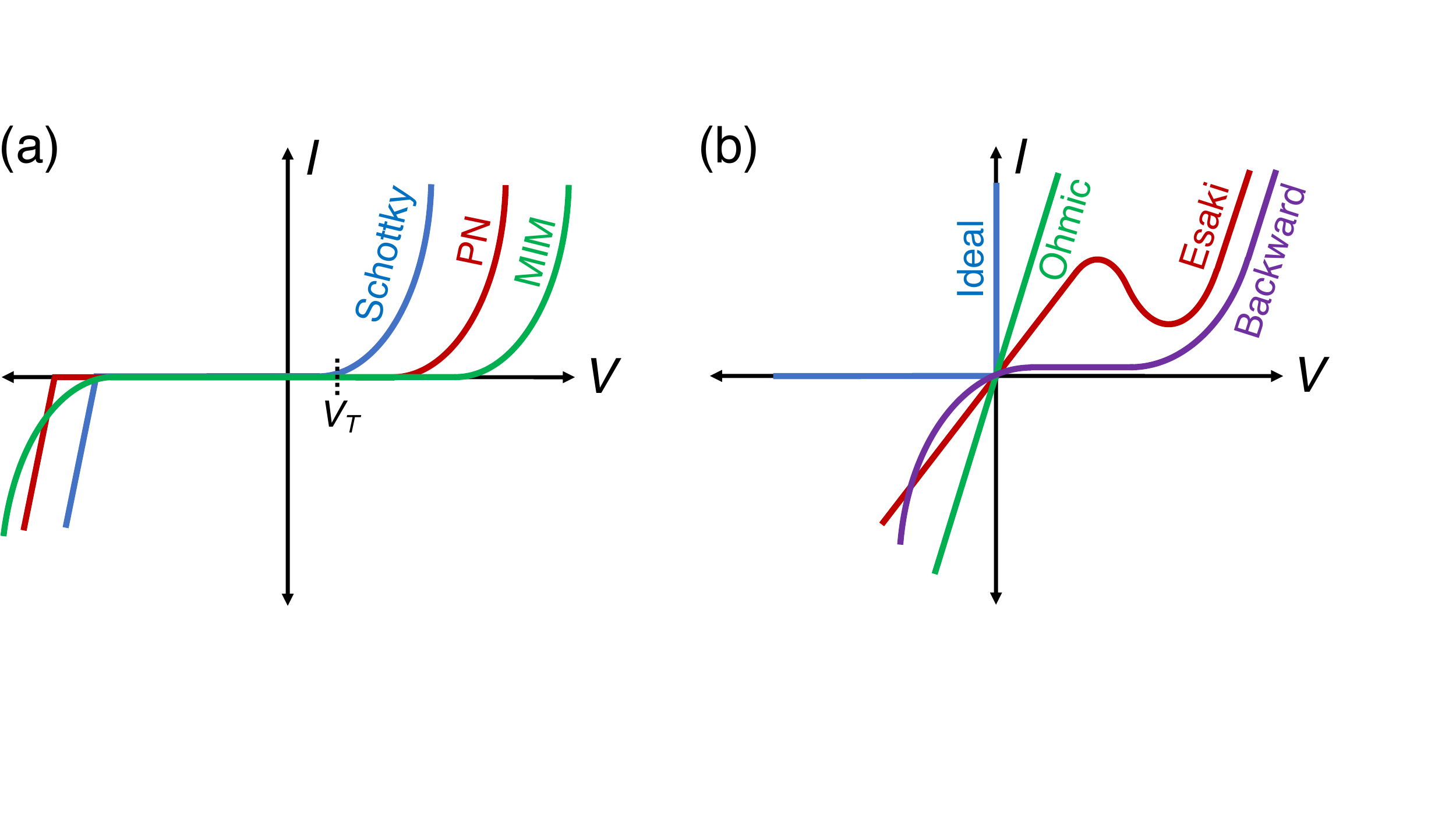}
\end{center}
\vspace{-1.8cm}
\caption{(a) A schematic representation of the current-voltage (\textit{I-V}) characteristics of 
the conventional three types of diodes [Schottky, \textit{p-n}, and metal-insulator-metal (\textit{M-I-M}) diodes]. 
(b) The \textit{I-V} curves of an ideal diode, an Ohmic junction, an Esaki (tunnel diode), and a backward diode.}
\label{fig1}
\end{figure}

A diode is a two-terminal device that conducts electric current in only one direction but restricts current from flowing in the opposite direction, i.e., it acts as a one-way switch for current. Diodes are also known as rectifiers because they change alternating current into direct current. Diodes are of several types, with different properties depending on the materials that they consist of~\cite{book_1,book_2,book_3}. For instance, \textit{p-n}-junction 
diodes are formed by joining a \textit{p}-type semiconductor with \textit{n}-type semiconductor and they 
are the elementary building blocks of the three-terminal transistors. The Esaki diode (or 
tunnel diode)~\cite{esaki1958new} is a heavily doped \textit{p-n}-junction diode, in which 
the electron transport in the contact region is via quantum-mechanical tunneling under 
forward bias and it shows the negative-differential-resistance (NDR) effect (see Fig.~\ref{fig1}), 
which allows it to function as oscillator and amplifier. In connection with the Esaki 
diode, when the doping concentration on the \textit{p} side or \textit{n} side is nearly or not quite degenerate, 
the current in the reverse direction is much larger than in the forward direction and hence 
such a device is called backward diode \cite{book_1,liu2017modulation,murali2018gate}. In 
contrast to semiconductor-semiconductor diodes, a Schottky-barrier diode~\cite{book_1} is 
formed by joining a metal with a \textit{n}-type semiconductor. Compared to typical \textit{p-n} junctions, 
Schottky diodes have very fast switching times and higher current drive capability.

Diodes are used for various applications, ranging from power conversion to overvoltage 
protection and from signal detection and mixing to switches. Due to the junction barrier, 
diodes have a threshold (or turn on) voltage $V_T$, which must be supplied to the diode 
for it to conduct any considerable forward current. In Fig.~\ref{fig1}, we show the current-voltage (\textit{I-V}) characteristics of the different types of diodes and 
compare them with the concepts of the ideal diode and the Ohmic junction. For \textit{p-n}-junction (silicon) 
diodes the threshold voltage $V_T$ is around 0.7~V, while for Schottky diodes $V_T$ 
is between 0.2~V and 0.3~V. Backward diodes have zero threshold voltage but their on:off 
current ratios as well as their voltage-operation windows are rather small. The threshold 
voltage $V_T$ gives rise to the power dissipation ($P=V_T \times I$) in the diode  in the form of heat 
and hence it is an undesirable feature. The larger the value of $V_T$, the higher is the power dissipation.

Although the ideal diode is a theoretical concept, it has been suggested that superconductor-semiconductor 
junctions possess \textit{I-V} characteristics that are similar to 
the ideal one under a forward bias \cite{SCSC_1}. Initial experiments by McColl 
\textit{et al.} on a superconductor-semiconductor junction diode based on lead and 
\textit{p}-type GaAs have shown ideal-diode behavior for forward applied voltages less 
than the superconducting-energy-gap parameter $\Delta$. In this voltage window 
(a few millielectronvolts) the diode exhibits a high degree of nonlinearity in its \textit{I-V} characteristics 
\cite{SCSC_1}. Superconductor-semiconductor junctions have subsequently been studied 
as the most sensitive detectors and mixers of microwaves \cite{SCSC_2,SCSC_3,SCSC_4,SCSC_5,SCSC_6}. 
However, in contrast to the Schottky diode behavior, superconductor-semiconductor 
junctions possess symmetric \textit{I-V} curves with $I(-V)=-I(V)$, close to zero bias, giving 
rise to a relatively small on:off current ratio~\cite{SCSC_6}. Furthermore, the operation
temperature of such diodes is limited by the phase-transition temperature of superconductors,
which is far below room temperature. Another type of diode that has a zero threshold 
voltage $V_T$ is the so-called geometric diode, the operation principle of which relies on the 
geometric asymmetry of a conducting thin film \cite{zhu2013graphene,zhu2014high,auton2017terahertz,kang2018terahertz}. 
Geometric diodes are ultra-fast ballistic transport devices, where the critical dimension 
of the device is comparable to the mean-free path length of the electrons. It was, however, been
shown that geometric asymmetry of the diode alone cannot induce a current rectification and
thus in addition to geometry, nonlinearity (high-order many-body interactions) is required to 
realize the geometric diode~\cite{li2014non}. This also explains why the experimentally measured
current rectification in graphene-related geometric diodes is very low \cite{zhu2013graphene,zhu2014high}
as the non-linearity, i.e., electron-phonon interactions are rather weak in these systems.

\begin{figure}[t]
\begin{center}
    \includegraphics[scale=0.158]{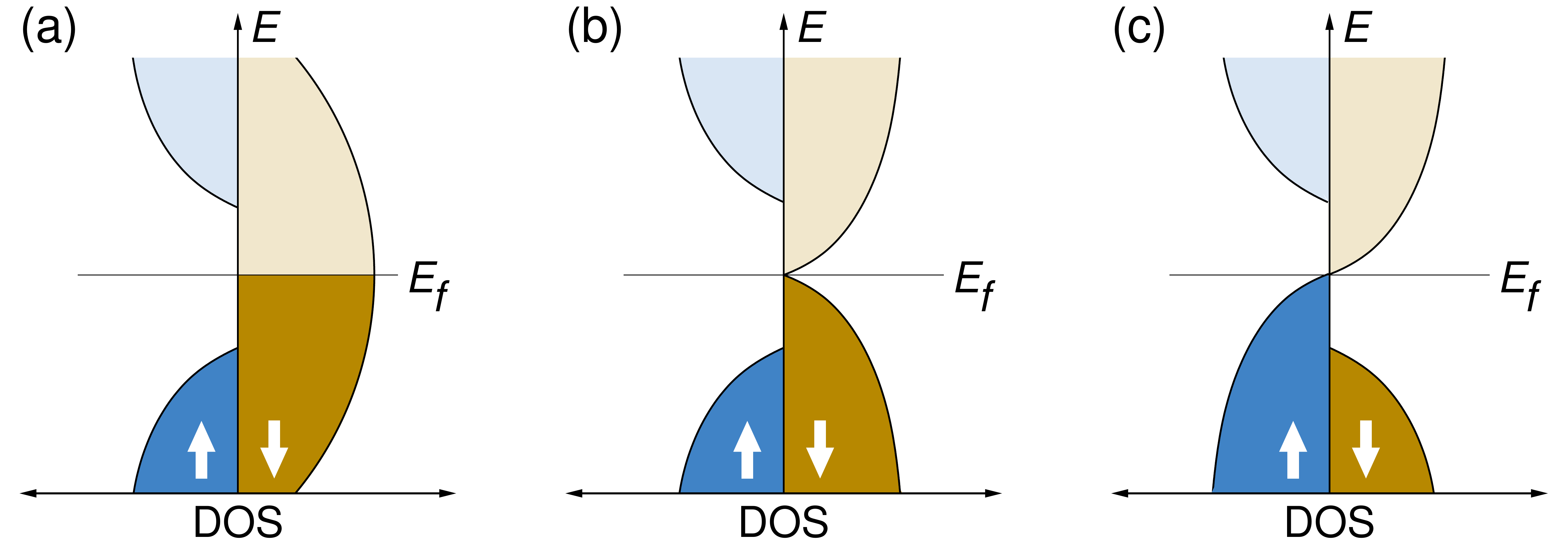}
\end{center}
\vspace{-0.3cm}
\caption{A schematic representation of the density of states
for (a) a half metallic magnet, (b) a type-I spin-gapless semiconductor, 
and (c) a type-II spin-gapless semiconductor.}
\label{fig2}
\end{figure}

For device applications, a special class of materials, the so-called spin-gapless 
semiconductors (SGSs), are receiving substantial attention. The concept of SGSs was proposed 
by Wang in 2008~\cite{wang2008,colossal}. By employing first-principles calculations 
the author predicted SGS behavior in the Co-doped dilute magnetic semiconductor PbPdO$_2$. 
Subsequently, different classes of materials ranging from two dimensional (2D) to three dimensional (3D) have been predicted to 
possess SGS behavior and some of them have been confirmed experimentally. The SGSs lie 
on the border between magnetic semiconductors and half-metallic magnets (HMMs)~\cite{de1983new}. 
A schematic density of states (DOS) of HMM and SGSs is shown in Fig~\ref{fig2}. In 
SGSs, the mobility of charge carriers is essentially higher than in normal semiconductors, 
making them promising materials for nanoelectronic applications. Moreover, the 
spin-dependent transport properties of SGSs and HMMs lead to the emergence of device
concepts in spintronics. Recently, a reconfigurable magnetic tunnel diode and transistor
concept based on SGSs and HMMs has been proposed~\cite{sasioglu2015patent, sasioglu2019proposal}. 
The magnetic tunnel diode allows electrical current to pass either in one or the other 
direction, depending on the relative orientation of the magnetization of the HMM and SGS 
electrodes. Moreover, the proposed devices present tunnel magnetoresistance effect, allowing 
the combination of nonvolatility and reconfigurability on the diode (transistor) level, 
which is not achievable in semiconductor nanoelectronics.

In this paper, we propose a diode concept, based on a HMM and a SGS electrode, that we call Ohmic spin diode (OSD) and demonstrate proof of the principle by \textit{ab-initio} quantum 
transport calculations. Analogous to the metal-semiconductor junction diode (the Schottky-barrier 
diode), HMM-SGS junctions act as a diode, the operation principle of which relies on the 
spin-dependent transport properties of the HMM and SGS materials. We show that HMM and SGS 
materials form an Ohmic contact under any finite forward bias, giving rise to linear 
current-voltage (\textit{I-V}) characteristics, while for a reverse bias the current is blocked 
due to the filtering of the electrons. In contrast to the Schottky diode, the proposed 
diode does not require the doping of the SGS and also does not have a junction barrier and 
thus it has a zero threshold voltage  $V_T$ and an infinite on:off current ratio at zero temperature. 
However, at finite temperatures non-spin-flip thermally excited high-energy electrons as well as
low-energy spin-flip excitations can give rise to leakage current and thus reduce the on:off 
ratio under a reverse bias. Moreover, due to the Ohmic HMM-SGS contact, the proposed diode has a 
much higher current drive capability  and low resistance, which is advantageous compared to 
conventional semiconductor diodes. To demonstrate the linear \textit{I-V} characteristics of the  
concept we construct a planar HMM-SGS junction based on 2D half-metallic Fe/MoS$_2$ 
and spin-gapless semiconducting VS$_2$ and employ the nonequilibrium Green’s function method combined 
with density-functional theory (DFT). We find that at zero bias the VS$_2$ and Fe/MoS$_2$ electrodes 
couple ferromagnetically; however. this coupling changes sign from ferro- to antiferromagnetic  
for a critical forward bias voltage of $V=180$~mV. The VS$_2$--Fe/MoS$_2$ junction diode possesses 
linear \textit{I-V} characteristics for forward bias voltages up to $V=180$~mV  and a very small threshold 
voltage of $V_T=30$~meV, which can be attributed to the minority-electron conduction-band minimum 
of the spin-gapless semiconducting VS$_2$ material. Moreover, we obtain a very high current density  
($J=2350$~$\mu$A/$\mu$m  for $V=180$~mV), which makes the VS$_2$--Fe/MoS$_2$ OSD highly promising 
for low-temperature nanoelectronic applications.

\section{HMM-SGS Junctions \label{sec:concept}}

\begin{figure}[t]
\centering
 \includegraphics[scale=0.34]{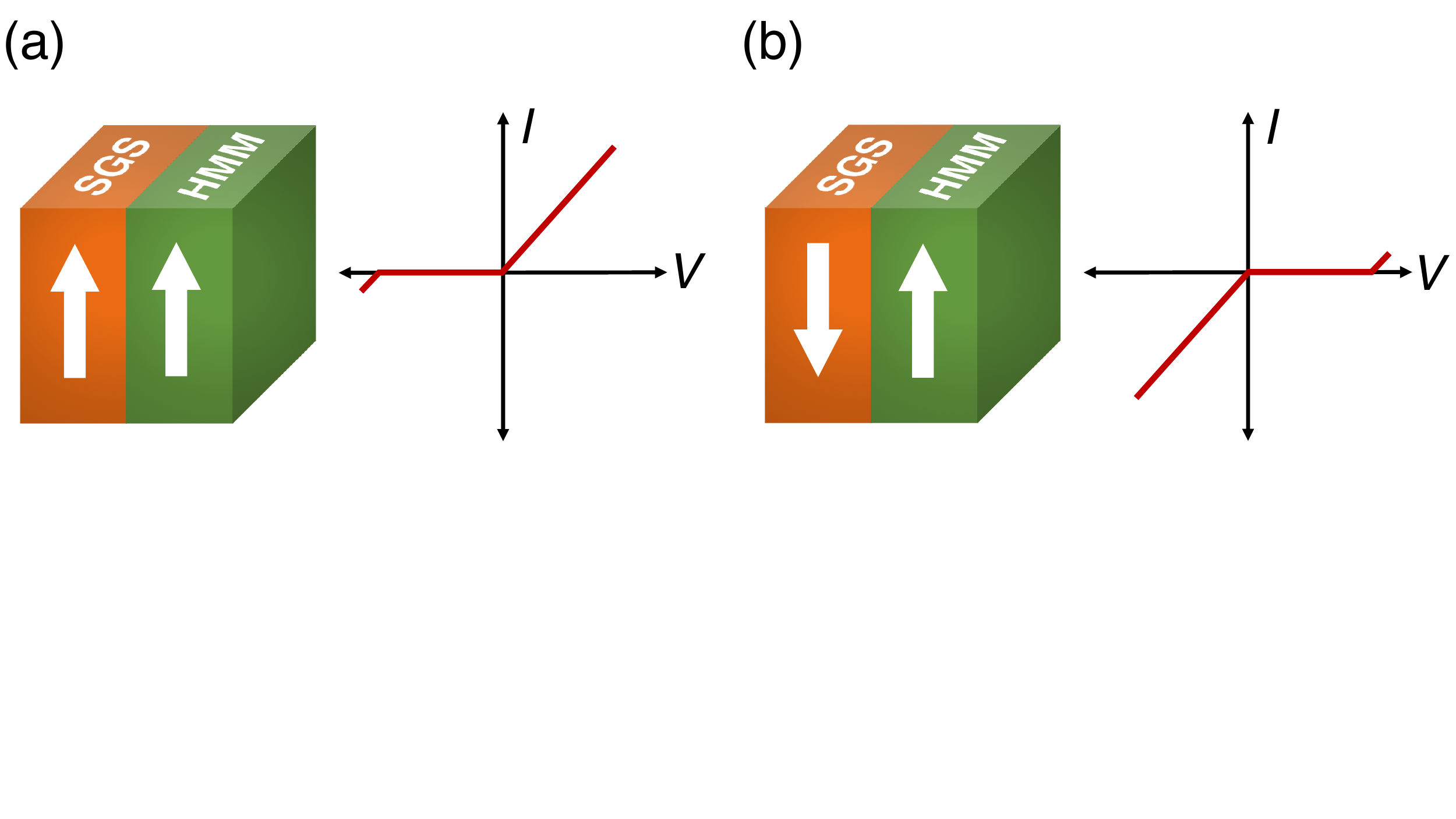}
\vspace{-2.6cm}
\caption{(a) A schematic representation of the HMM-SGS junction 
for parallel orientation of the magnetization directions of the electrodes 
and the corresponding current-voltage (\textit{I-V}) curves. (b) The same as (a) for antiparallel 
orientation of the magnetization directions of the electrodes. The white arrows 
indicate the direction of the magnetization of the electrodes.}
\label{fig3}
\end{figure}

The structure of the proposed OSD and its \textit{I-V} characteristics are shown schematically 
in Fig.~\ref{fig3}. Analogous to the metal-semiconductor Schottky-barrier 
diode, the OSD is composed of a HMM electrode and a type-II SGS electrode. Depending 
on the magnetization direction of the electrodes, the diode conducts current either 
under forward bias [Fig.~\hyperref[fig3]{3(a)}, parallel orientation (ferromagnetic 
interelectrode coupling)] or under the reverse bias  [Fig.~\hyperref[fig3]{3(a)}, antiparallel 
orientation (antiferromagnetic interelectrode coupling)], similar to the case of the backward 
diode mentioned in Sec.~\ref{sec:Intro}. In the presentation of the schematic \textit{I-V} characteristics 
of the OSD in Fig.~\ref{fig3}, we assume that HMM material possesses a gap in the 
spin-up channel around the Fermi level, while type-II SGS material has a gap in the 
spin-up (spin-down) channel above (below) the Fermi level, as shown schematically in 
Fig.~\ref{fig2}. In type-II SGSs, the conduction- and valence-band edges of the different 
spin channels touch at the Fermi energy, while in type-I SGSs the spin-up (majority-spin)
band looks like the on in HMMs but the difference is in the spin-down (minority-spin) band. The 
valence- and conduction-band edges are touching at the Fermi energy, so that a zero-width gap exists.
One of the important advantages of type-I SGSs is that no energy is required 
for the excitation of the electrons from the valence to the conduction band and excited 
electrons or holes can be 100\% spin-polarized like in HMMs. In construction of the OSD, 
the useo possible. of a type-I SGS instead of the HMM is als

\begin{figure*}[bt]
\begin{center}
 \includegraphics[scale=0.167]{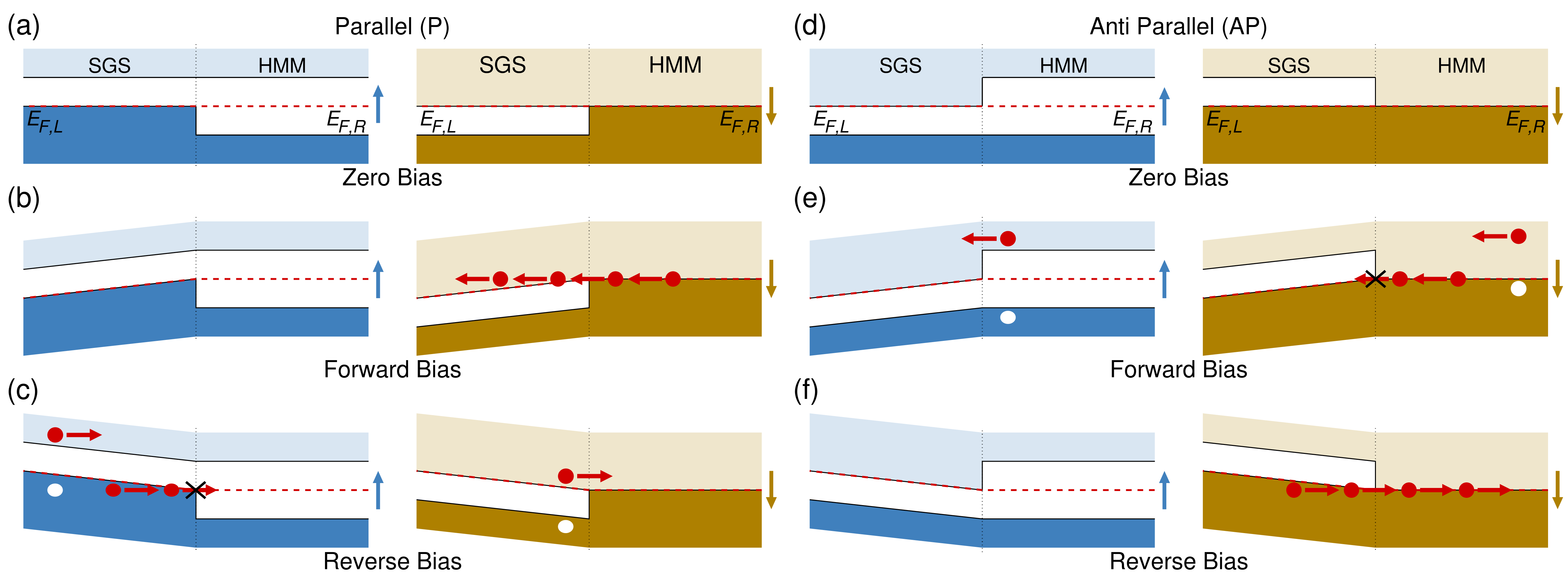}
\end{center}
\vspace{-0.5 cm}
\caption{A schematic representation of the band diagram and thermal
(non-spin-flip) electron-hole excitations for spin-up and spin-down electrons for 
the SGS-HMM contact for parallel (P) orientation of the magnetization directions of 
the  electrodes: (a) for zero bias, (b) for forward bias, and (c) for reverse bias. 
The Fermi level is denoted by red dashed line. (d)--(f) The same as (a)--(c) 
for the antiparallel (AP) orientations of the magnetization directions 
of the  electrodes [see Figs.~\hyperref[fig3]{3(a)} and \hyperref[fig3]{3(b)}].}
\label{fig4}
\end{figure*}

The linear \textit{I-V} characteristics of the OSD presented in Fig.~\hyperref[fig3]{3(a)} can be 
qualitatively explained on the basis of the schematic energy-band diagram shown in Fig.~\ref{fig4}. 
If we assume that both HMM and SGS electrodes have the same work function and equal Fermi levels, 
then no charge transfer takes place between the electrodes. However, in real materials
due to different work functions there might be a charge transfer from one material to another  
at the interface, which might give rise to band bending for the SGS. Besides this, due to
interactions, junction materials might not retain their half metallic and spin-gapless semiconducting
properties near the interface and hence the band diagram would not be as sharp as in Fig.~\ref{fig4}. 
For the device configuration shown in Fig.~\hyperref[fig3]{3(a)} the relevant channel for the transport 
is the minority-spin (spin-down) channel, whereas the majority-spin (spin-up) channel is insulating 
due to the spin gap of HMM material on the right-hand side of the junction. In the spin-down channel, the HMM 
behaves like a normal metal with states below and above the Fermi energy, while the SGS electrode 
on the left-hand side behaves like a semiconductor (or insulator) but with a Fermi level touching the 
conduction-band minimum. Due to this electronic band structure of the SGS, in contrast to 
the Schottky diode, no energy barrier is formed at the interface between a HMM and a SGS material. 
Such a junction acts as an Ohmic contact under forward bias as shown in Fig.~\hyperref[fig4]{4(b)}. In 
this case, the spin-down electrons from the occupied valence band of the HMM electrode can flow into 
the unoccupied conduction band of the SGS electrode without experiencing a potential barrier, while 
for a reverse bias the current is blocked due to the spin gap of HMM material [see Fig.~\hyperref[fig4]{4(c)}].
Note that under a forward bias, the current flowing through the OSD is 100\% spin polarized. Note also
that the same discussion applies in the case of  antiparallel orientations of the magnetization directions 
of the electrodes [see Fig.~\hyperref[fig4]{4(d)}, \hyperref[fig4]{4(e)}, and \hyperref[fig4]{4(f)}].

As the HMM-SGS contact is Ohmic under a forward bias the current $I$ through the diode varies 
linearly with the applied voltage $V$ and the ratio $V/I$ gives the combination of the interface 
($R_I$) and series resistance ($R_S$) of the HMM and SGS materials ($V/I=R_I+R_S$). The resistance 
of SGS materials is usually much lower than that of conventional \textit{n}- or \textit{p}-type semiconductors~\cite{ouardi2013realization} 
and thus the combination of low resistance with the Ohmic nature of the 
interface allows a much higher current drive capability of the proposed OSD. It is worth noting 
that diodes with low resistance are critical for the performance of high-speed electronic devices. 
Besides the higher current drive capability of the OSD, the threshold voltage $V_T$ can be tuned by a 
proper choice of the SGS material. The value of $V_T$ is set by the energy difference between the minority-spin 
conduction-band minimum and the Fermi level. In an ideal SGS, this difference is zero and thus 
$V_T=0$. Note, however, that in type-II SGSs, the spin-gapless semiconducting properties are not 
protected by any symmetry and thus ideal SGS behavior can only arise if a free parameter -- e.g. pressure, 
strain, or doping -- is tuned to a specific value. A zero $V_T$ allows one to detect extremely weak signals, 
even when no external bias circuit is used. Similar to the Schottky diode, the OSD is also a majority-carrier 
diode but it does not require doping and it will possess all the advantages of the Schottky diode, such as 
high operation frequencies, low resistance and capacitance, etc. Note also that, in principle, a Schottky diode 
can be turned into an Ohmic contact by heavily doping the semiconductor electrode; 
however, in this case it loses its diode functionality. 

\begin{figure}[b]
\begin{center}
    \includegraphics[width=\columnwidth]{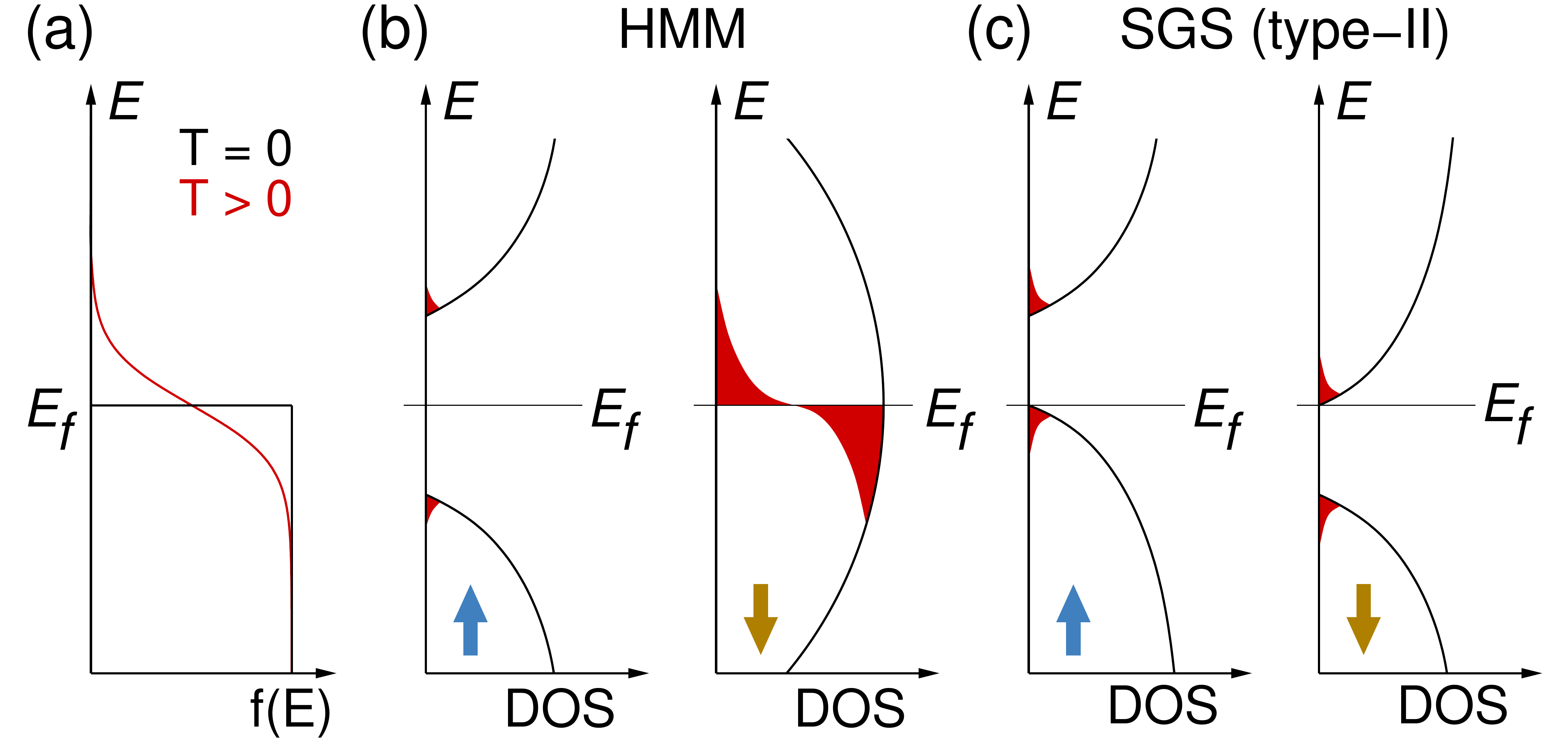}
\end{center}
\vspace{-0.3cm}
\caption{(a) The Fermi-Dirac distribution for $T = 0$~K (black) and $T>0$~K (red)
and thermal population of states around the Fermi level for (b) a HMM and (c) type-II SGS. The 
occupied and unoccupied states above and below the Fermi level, respectively, are marked with a shaded red color.}
\label{fig5}
\end{figure}

Up to now, the discussion of the \textit{I-V} characteristics of the OSD is based on the schematic band 
diagram at zero temperature and thus an infinite on:off current ratio can be achieved. However, 
at finite temperatures, thermally excited electrons and holes can significantly reduce the on:off 
ratio. At this point, it is important to note that in conventional metal-semiconductor Schottky diodes 
electrons flow from semiconductor to the metal electrode under a forward bias, whereas in the OSD 
shown in Fig.~\hyperref[fig3]{3(a)} the process is just the opposite, i.e, from the half-metal to the spin-gapless 
semiconductor. In the former case, the on-state current stems from the combination of thermionic emission over 
the Schottky barrier and tunnelling through the barrier~\cite{stradi2016general,zhang2006analysis,latreche2019combined}. 
A very detailed  analysis of these processes and their relative contributions to the total forward (on-state) 
current in Schottky diodes can be found in Ref.~\onlinecite{stradi2016general}. Similarly, under a reverse bias, 
both processes contribute to the leakage current in Schottky diodes, with a significant weight coming from 
the tunneling (or thermionic field emission), since in this case the height of the Schottky barrier is 
fixed and only a very small fraction of the thermally excited high-energy electrons, which are at the tail 
of the Fermi-Dirac distribution function, can pass over the barrier. By increasing the height of the Schottky 
barrier, the contribution of thermionic emission can be significantly reduced, as the Fermi-Dirac distribution 
function decays exponentially at high energies [see Fig.~\hyperref[fig3]{3(a)}]; however, the contribution 
of the tunneling through the barrier cannot be prevented and at the same time the threshold voltage $V_T$
increases. In the OSD, the leakage current due to tunneling does not exist; however, thermally excited 
high-energy electrons contribute to the reverse current, similar to the case of Schottky diode. 
In Figs.~\hyperref[fig5]{5(b)} and \hyperref[fig5]{5(c)}, we show schematically the population of unoccupied 
states around the Fermi energy for a HMM and a type-II SGS. In a HMM, thermal population of the states obeys the 
Fermi-Dirac distribution, i.e., more states in the metallic spin-channel and very few states for the 
insulating spin-channel, as they are far from the Fermi level. However, the situation turns out to be
slightly different in type-II SGSs, due to their band structure, i.e., only a small fraction of the
electrons can be thermally excited due to the gap in both spin channels, somewhat similar to the case of
intrinsic semiconductors. Consequently, the reverse bias current (or the leakage current) is mainly
determined by the size of the band gaps in SGS and HMM electrodes and by a proper choice of the large
band-gap materials, the on:off current ratio can be significantly increased in OSDs. In addition to
thermal excitations, spin-flip processes can reduce the on:off current ratio and this will be discussed
later.

\begin{table}[!t]
\caption{The \textit{I-V} characteristics of the OSD for parallel (P) and antiparallel (AP) 
orientations of the magnetization directions of the electrodes and all possible 
combinations of the spin character of the gaps in HMMs and SGSs. SGS\,($\downarrow \uparrow$) 
indicates the spin channel where the gap exists below and above the Fermi level.}
\centering
\begin{ruledtabular}
\begin{tabular}{@{}*{4}{c}@{}}
     Spin gap & Orientation & Forward bias & Reverse bias\\
\hline
SGS\,($\downarrow \uparrow$)/HMM\,($\uparrow \uparrow$)     & P  & On & Off \\
SGS\,($\downarrow \uparrow$)/HMM\,($\uparrow \uparrow$)     & AP & Off & On \\
SGS\,($\downarrow \uparrow$)/HMM\,($\downarrow \downarrow$) & P  & Off & On  \\
SGS\,($\downarrow \uparrow$)/HMM\,($\downarrow \downarrow$) & AP & On & Off \\
SGS\,($\uparrow \downarrow$)/HMM\,($\uparrow \uparrow$)     & P  & Off & On \\
SGS\,($\uparrow \downarrow$)/HMM\,($\uparrow \uparrow$)     & AP & On & Off \\
SGS\,($\uparrow \downarrow$)/HMM\,($\downarrow \downarrow$) & P  & On & Off \\
SGS\,($\uparrow \downarrow$)/HMM\,($\downarrow \downarrow$) & AP & Off & On \\
\end{tabular}
\end{ruledtabular}
\label{tab1}
\end{table}

In contrast to the Schottky diode, the spin degree of freedom provides a rich configuration space 
for the \textit{I-V} curves of the OSD, which are determined by two parameters: (i) the magnetic coupling 
between electrodes, which allows the dynamical configuration of the diode in the case of 
antiferromagnetic interelectrode coupling via an external magnetic field; and (ii) the spin character 
of the gap in HMMs and SGSs. Depending on the magnetic coupling between electrodes, the OSD is in the on state 
either under forward bias (ferromagnetic interelectrode coupling or parallel orientation) or 
under reverse bias (antiferromagnetic interelectrode coupling or anti-parallel orientation), as 
shown in Fig.~\ref{fig3}. The second parameter, which plays a decisive role in determining the \textit{I-V} 
characteristics of the OSD, is the spin-channel dependence of the gap in HMM and SGS electrodes. 
For instance, as mentioned above in the presentation of the schematic \textit{I-V} characteristics of the 
OSD in Fig.~\ref{fig3}, we assume that the HMM material has a gap in the spin-up channel and the SGS material has 
gaps in the spin-down (below $E_F$) and spin-up channels (above $E_F$). Although all known type-II 
SGSs possess an electronic band structure similar to that in Fig.~\hyperref[fig2]{2(c)}, there 
are many HMMs, such as Heusler alloys, with a gap in the spin-down channel and OSD diodes constructed 
from such materials might have a different current direction than the present case. For completeness, 
in Table~\ref{tab1} we present the \textit{I-V} characteristics of the OSD by taking into account both 
magnetic configurations of the electrodes and all possible combinations of the spin character 
of the gaps in HMMs and SGSs.

As the OSD is comprised of magnetic materials, its operation temperature is limited by the Curie 
temperature $T_C$ of the constituent materials and thus for realization of the OSD, HMMs and SGSs 
with high $T_C$ values are required. Two-dimensional transition-metal dichalcogenides \cite{xu2013graphene,mounet2018two,haastrup2018computational,gibertini2019magnetic}  
and 3D quaternary Heusler compounds \cite{ozdougan2013slater,TMR_H_3,graf2011simple,gao2019high,aull2019ab,rotjanapittayakul2018search} 
offer a unique platform to design, within the same family of compounds, HMMs and SGSs with high $T_C$ 
values and similar lattice parameters and compositions, which allow coherent growth of these materials 
on top of each other. Besides high $T_C$ values, large spin gaps in HMMs and SGSs are crucial to achieve 
high on:off current ratio in the OSD. In recent years, 2D transition-metal dichalcogenides have received 
significant experimental and theoretical interest, as they present unique electronic, optical, mechanical, 
and magnetic properties, thus holding great promise for a wide range device applications. Devices ranging 
from vertical tunnel diodes to vertical and lateral tunnel field-effect transistors (TFETs) have been 
experimentally demonstrated \cite{radisavljevic2011single,li2014black,jariwala2014emerging,fiori2014electronics,roy2015dual,liu2016van,yan2015esaki,giannazzo2018vertical}. 
In particular, 2D lateral heterojunctions have opened up a direction in materials science and device 
applications~\cite{li2016heterostructures}. TFETs based on 2D material heterojunctions (WTe$_2$-MoS$_2$, 
MoTe$_2$-MoS$_2$) have been reported to exhibit subthreshold slope below 5~mV/dec and high 
$I_{\rm{ON}}/I_{\rm{OFF}}$ ratios (approximately $10^8$) at a low drain bias of 0.3~V, 
making them ideal candidates for ultralow-power computing~\cite{choukroun2018high}.

Among the 2D transition-metal dichalcogenides, \textit{V}-based compounds (VS$_2$, VSe$_2$, VTe$_2$ ) have attracted 
particular interest due to their intrinsic ferromagnetism. These compounds can crystallize in two different 
structures: the 1\textit{H} phase  and the 1\textit{T} phase. The former phase is energetically more stable and possesses a SGS 
ground state \cite{zhuang2016stability,luo2017structural}. Although, theoretically, both 1\textit{H} and 1\textit{T} phases of
\textit{V}-based 2D compounds have been predicted to show ferromagnetism~\cite{fuh2016newtype}, experimentally the 
ferromagnetism has, however, only been observed in the 1\textit{T} phase of VS$_2$~\cite{gao2013ferromagnetism}  
and VSe$_2$~\cite{bonilla2018strong}. Note that the 1\textit{T} phase of VSe$_2$ does not present SGS behavior; 
it is a normal ferromagnetic metal. However, in the one-monolayer limit, several 2D transition-metal dichalcogenides can adopt either a 1\textit{T} or a 1\textit{H} structure depending on the growth conditions~\cite{han2018van}.  It is very 
likely that VS$_2$ can also be grown in a 1\textit{H} structure.

\section{Computational Method \label{sec:Method}}

Ground state electronic structure calculations are carried out using density functional 
theory (DFT), implemented in the QuantumATK P-2019.03 package~\cite{QuantumATK}. We use the
generalized-gradient-approximation (GGA)--Perdew-Burke-Ernzerhof (PBE) exchange-correlation functional~\cite{perdew1996generalized} together with PseudoDojo 
pseudopotentials~\cite{QuantumATKPseudoDojo} and LCAO basis-sets. A dense 
$24 \times 24 \times 1$ $\mathbf{k}$-point grid and density mesh cutoff of 120~hartree are used. To prevent interactions between the periodically repeated images, 20~{\AA} of vacuum 
are added and Dirichlet and Neumann boundary conditions are employed. The electron temperature
is set to 10~K. The total energy and forces converge to at least to 10$^{-4}$~eV and 
0.01~eV/{\AA}, respectively. In order to estimate magnetic anisotropy energy, we employ the 
magnetic force theorem, including spin-orbit coupling~\cite{smidstrup2019an}.

The transport calculations are performed using DFT combined with the nonequilibrium 
Green's function method (NEGF). We use a $24 \times 1 \times 172$ $\mathbf{k}$-point 
grid in self-consistent DFT-NEGF calculations. The \textit{I-V} characteristics were calculated 
within a Landauer approach~\cite{Landauer-Buettiker}, where $ I(V) = \frac{2e}{h}\sum_{\sigma}\int \, 
T^{\sigma}(E,V)\left[f_{L}(E,V)-f_{R}(E,V)\right] \mathrm{d}E $. Here, $V$ denotes the bias voltage, 
$T^\sigma (E,V)$ is the spin-dependent transmission coefficient for an electron with 
spin $\sigma$, and $f_L(E,V)$ and $f_R(E,V)$ are the Fermi-Dirac distributions of the left 
and right electrodes, respectively. The transmission coefficient $T^\sigma (E,V)$ is 
calculated using a $300 \times 1$ $\mathbf{k}$-point grid.

\section{Results and Discussion \label{sec:ResandDis}}

\begin{table}[!b]
\caption{The calculated lattice parameter $a$, total magnetic moment $m_T$, magnetic anisotropy 
energy $K$ (per formula unit), and the work-function $\phi$ for VS$_2$ and Fe/MoS$_2$. The values of the Curie temperature 
T$_c$ are taken from the literature.}
\centering
\begin{ruledtabular}
\begin{tabular}{@{}l*{5}{c}@{}}
 Compound & $a$ & $m_T$ & K & $\phi$ & T$_C$ \\
          & ({\AA}) & ($\mu_B$) & (meV) & (eV) & (K) \\
  \hline
VS$_2$ & 3.174     & 1.00 & 0.20 & 5.71 & 138~\cite{fuh2016newtype} \\
Fe/MoS$_2$ & 3.175 & 2.00 & 0.42 & 4.72 & 465~\cite{jiang2018robust}\\
\end{tabular}
\end{ruledtabular}
\label{tab2}
\end{table}

\begin{figure*}[!t]
\begin{center}
\includegraphics[width=0.95\textwidth]{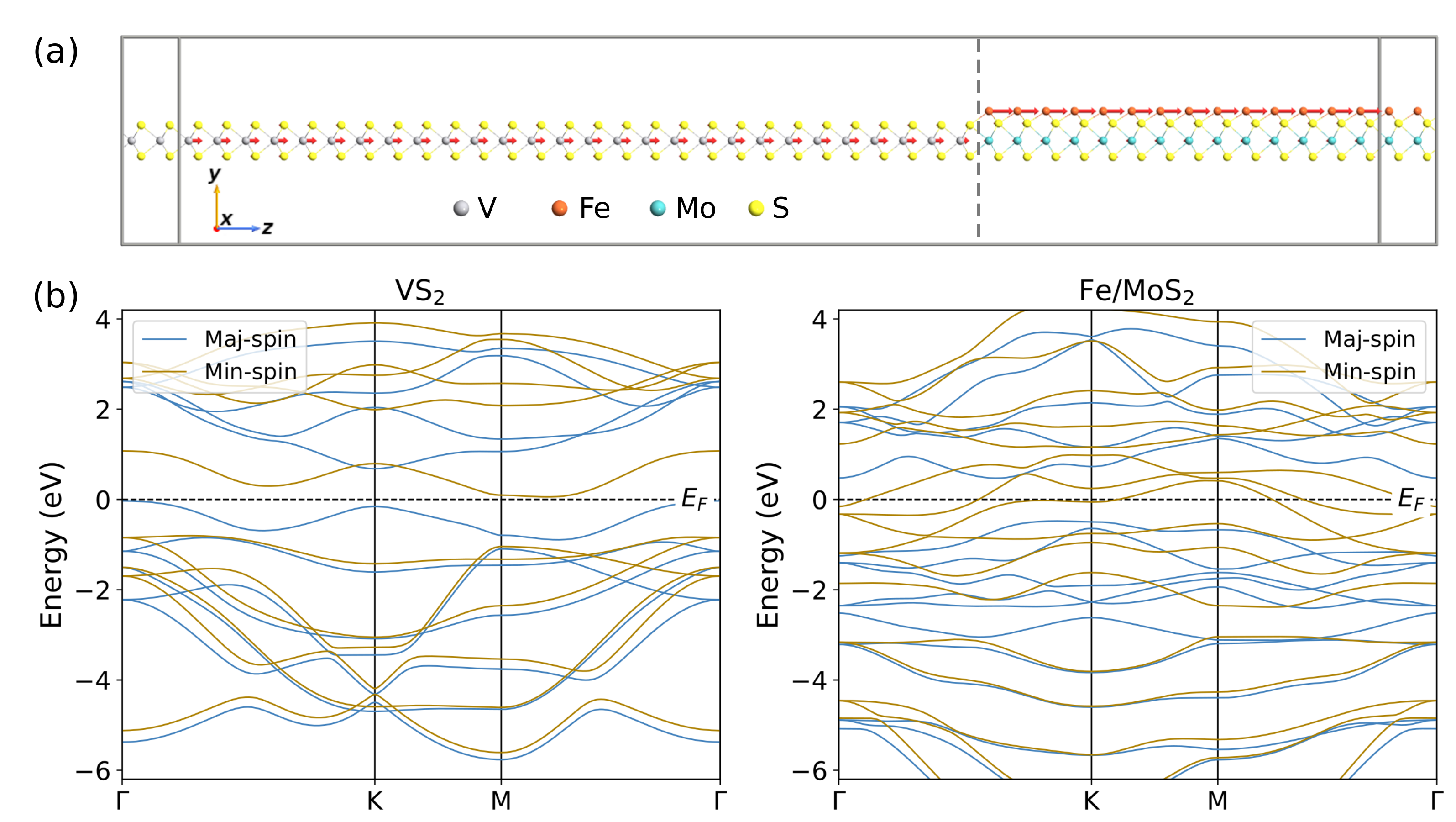}
\end{center}
\vspace*{-0.6cm} 
\caption{(a) The atomic structure of the VS$_2$-Fe/MoS$_2$ Ohmic spin diode. The arrows 
show the magnetic moments of V and Fe atoms in the SGS and HMM electrodes within the 
scattering region. The size of the arrows is proportional to the magnitude of the
magnetic moments. The system is periodic in the $x$-direction in the plane orthogonal 
to the magnetization direction, which is also the transport direction. The vertikal dashed lines denote the interface. (b) The calculated 
spin-resolved bulk electronic band structure along the high-symmetry lines in the 
Brillouin zone for VS$_2$ (left-hand panel) and Fe/MoS$_2$ (right-hand panel). For both compounds,  
the dashed lines denote the Fermi energy, which is set to zero.} 
\label{fig6}
\end{figure*}

The OSD concept introduced in Sec.~\ref{sec:concept} can be realized either by using 3D Heusler 
compounds or 2D transition-metal dichalcogenides. In the following, due to their structural 
simplicity we focus on the 2D materials and demonstrate the proof of principle by \textit{ab-initio} 
quantum transport calculations. As the electrode materials of the OSD, we choose VS$_2$ and 
Fe/MoS$_2$. The former is an intrinsic ferromagnet with spin-gapless semiconducting behavior 
in the monolayer 1\textit{H} phase, while the half-metallic ferromagnetism in the latter 
Fe/MoS$_2$ material is achieved by functionalization of the 1\textit{H} semiconducting MoS$_2$. 
Based on first-principles calculations, Jiang \textit{et al}.~\cite{jiang2018robust} have shown that deposition of the Fe atoms 
on MoS$_2$ gives rise to the 2D half-metallic ferromagnetism with a relatively high $T_C$ 
value of 465~K and a large spin gap, which makes the Fe/MoS$_2$ a promising material for 
spintronic and nanoelectronic applications.

Since the electronic and magnetic as well as structural properties of both electrode materials 
have been extensively discussed in the literature, in the following we will briefly overview 
their basic properties, which will be necessary in order to understand the transport characteristics 
of the OSD. In Table~\ref{tab2}, we present the optimized lattice constants and the total magnetic 
moments as well as the magnetic anisotropy energies for the 1$H$ phase of VS$_2$ and Fe/MoS$_2$. 
Our ground-state calculations for both materials are in good agreement with previous published 
data. In particular, similar lattice parameters and compositions, as well as the same 1$H$ phase, 
of the two materials are crucial for realization of the planar VS$_2$-Fe/MoS$_2$ heterojunctions. 
VS$_2$ has a relatively simple band structure, shown in Fig.~\hyperref[fig6]{6(b)}, where the exchange 
splitting of the V-3$d$ (predominantly $d_{z^{2}}$) bands around the Fermi energy is responsible 
for its spin-gapless semiconducting nature and thus it has a magnetic moment of 1~$\mu_B$, carried 
by the V atom. Furthermore, VS$_2$ is not a perfect SGS; it has a very small indirect band gap 
of 50~meV, i.e., the valence-band maximum and conduction-band minimum are at around 20~meV and 30~meV, 
respectively.  The former plays a decisive role in determining the threshold voltage $V_T$ of the 
OSD in the case of ferromagnetic coupling of the VS$_2$ and Fe/MoS$_2$ electrodes, while the latter 
plays the same role in the case of antiferromagnetic coupling. Note that in the present OSD based on 
a planar VS$_2$-Fe/MoS$_2$ heterostructure, the coupling between the electrodes is ferromagnetic, as 
is discussed later. On the other hand, the Fe-deposited MoS$_2$ turns into a half-metallic
magnet with a gap of about 1~eV in the spin-up channel and with a total magnetic moment of 2~$\mu_B$, 
which is localized on the Fe atom. Note that PBE is well know to underestimate the band gap of 
semiconductors compared to the more accurate $GW$ approach. However, the situation is different 
for 2D SGSs, since the application of $GW$ method for the similar material VSe$_2$ reduces the band 
gap from 250 meV (PBE) to 170 meV~\cite{haastrup2018computational}. We expect a similar behavior
for the VS$_2$ compound when the $GW$ method is employed. Also, spin-orbit coupling has a negligible effect 
on the spin polarization of both materials and thus it is not taken into account in our device calculations .

In Fig.~\hyperref[fig6]{6(a)}, we show the atomic structure of the OSD, which is formed by joining 
one monolayer of VS$_2$ (left electrode) and one monolayer of Fe/MoS$_2$ (right electrode) laterally 
in a single plane. Due to the almost identical lattice parameters of both materials, as well as 
their similar compositions, they form a perfect interface. We assume periodicity of the device in the 
$x$-direction. The $z$-direction is chosen as the transport direction. The total length of the 
scattering region is about 115.5~{\AA}, which consists of 77~{\AA} VS$_2$ and 38.5~{\AA} Fe/MoS$_2$. 
The length of the former electrode is chosen larger because of the longer screening length in SGSs.

\begin{SCfigure*}
\includegraphics[width=0.75\textwidth]{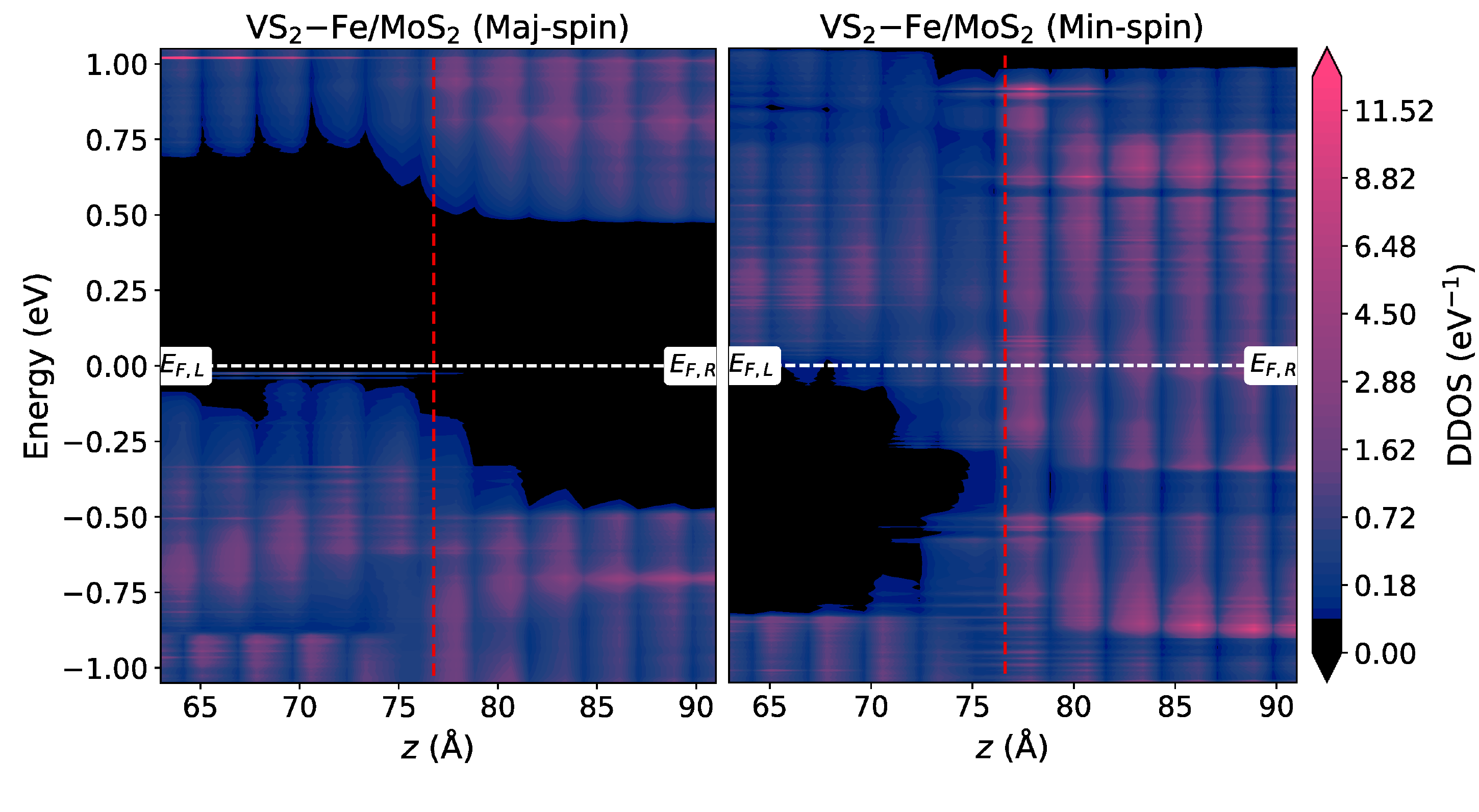}
\caption{The zero-bias projected device density of states (DDOS) for the majority (left hand panel) 
and minority (right hand panel) electrons for the Ohmic spin diode around the interface (see Fig.~\ref{fig6} 
for the atomic structure of the OSD). The horizontal white dashed lines indicate the Fermi level. The vertical red 
dashed lines denote the interface.}
\label{fig7}
\end{SCfigure*}

When the half metallic Fe/MoS$_2$ makes contact with VS$_2$, free electrons will flow from the half-metallic 
Fe/MoS$_2$ side to the spin-gapless semiconducting VS$_2$ side, because the work function of 
the Fe/MoS$_2$ is smaller than that of VS$_2$ (see Table~\ref{tab2}). Note that the work function of the 
SGS VS$_2$ is defined in the same way as in metals, i.e., the energy difference between the vacuum level 
and the Fermi energy $E_F$. When the charge redistribution reaches the equilibrium, near the interface 
region the Fe/MoS$_2$ will be positively charged, whereas the VS$_2$ will be negatively charged. Thus, 
an electric dipole will be induced at the interface region. Such a charge redistribution influences the 
electronic and magnetic properties of the materials near the interface, as seen in the zero-bias projected 
device density of states (DDOS) shown in Fig.~\ref{fig7}. Near the interface, charge transfer takes 
place within the spin-down channel and thus the magnetic moment of the Fe atoms increases towards the interface, 
i.e., 2.15~$\mu_B$ $\rightarrow$ 2.17~$\mu_B$ $\rightarrow$ 2.56~$\mu_B$, as shown by the arrows in 
Fig.~\hyperref[fig6]{6(a)}, where the size of the arrows is proportional to the magnitude of the magnetic 
moments. The transferred charge occupies the spin-down channel on the VS$_2$ side by creating interface 
states, which can be clearly seen on the projected DDOS shown in Fig.~\ref{fig7} and as a consequence 
the magnetic moment of the V atoms at the interface and subinterface lines is reduced from its bulk 
value of 1.13~$\mu_B$ to 0.53~$\mu_B$ and 0.95~$\mu_B$, respectively. As can be seen, the influenced region 
is rather small, being within four atomic lines and restricted to the spin-down channel. The change in the 
spin-up channel is more or less negligible. 

Long-range magnetic order is prohibited in the 2D magnets at finite temperatures due to the Mermin-Wagner
theorem~\cite{mermin1966absence}. However, this restriction can be removed by magnetic anisotropy and it 
enables the occurrence of 2D magnetic order at finite temperatures. For both materials, we have calculated 
magnetic anisotropy energies $K$, which are presented in Table~\ref{tab2}. Both materials have an in-plane 
magnetization with $K$ values of 0.2~meV (VS$_2$) and 0.42~meV (Fe/MoS$_2$). However, magnetic 
anisotropy within the plane for VS$_2$ is negligibly small (a few nanoelectronvolts), while for Fe/MoS$_2$ it is around 5~$\mu$eV, 
which is large enough for a finite-temperature magnetic order. The negligible value of $K$ for 
VS$_2$ implies the lack of finite-temperature magnetic order by virtue of the Mermin-Wagner theorem. However, 
room-temperature ferromagnetism has been experimentally detected in similar materials such as VSe$_2$ \cite{bonilla2018strong}, 
which has also negligible in-plane magnetic anisotropy, and the origin of the long-range ferromagnetic order 
is attributed to the finite-size effects~\cite{bramwell1994magnetization,torelli2019high}. However, 
in VS$_2$-Fe/MoS$_2$ junction, the Fe/MoS$_2$ acts as a pinning electrode, which introduces a 
preferred in-plane magnetic orientation in the VS$_2$ electrode via the ferromagnetic interelectrode
coupling. Our calculations show that ferromagnetic interelectrode coupling is preferable compared to 
antiferromagnetic interelectrode coupling and that the energy difference between the two configurations is about 22 meV.

Besides the magnetic anisotropy energy $K$ and the interelectrode coupling, another important parameter 
for the realization of the OSD is the Curie temperature $T_C$ of the constituent materials. The 
$T_C$ values for both materials have been estimated from first principles in 
Refs~\citenum{fuh2016newtype} and \citenum{jiang2018robust} (see Table~\ref{tab2}). The $T_C$ 
value of Fe/MoS$_2$ is higher than room temperature, while for VS$_2$ ($T_C=139$~K) is below 
room temperature. Nevertheless, such a value is high enough for an experimental demonstration of 
the device.

\begin{figure*}[!ht]
\begin{center}
\includegraphics[width=0.98\textwidth]{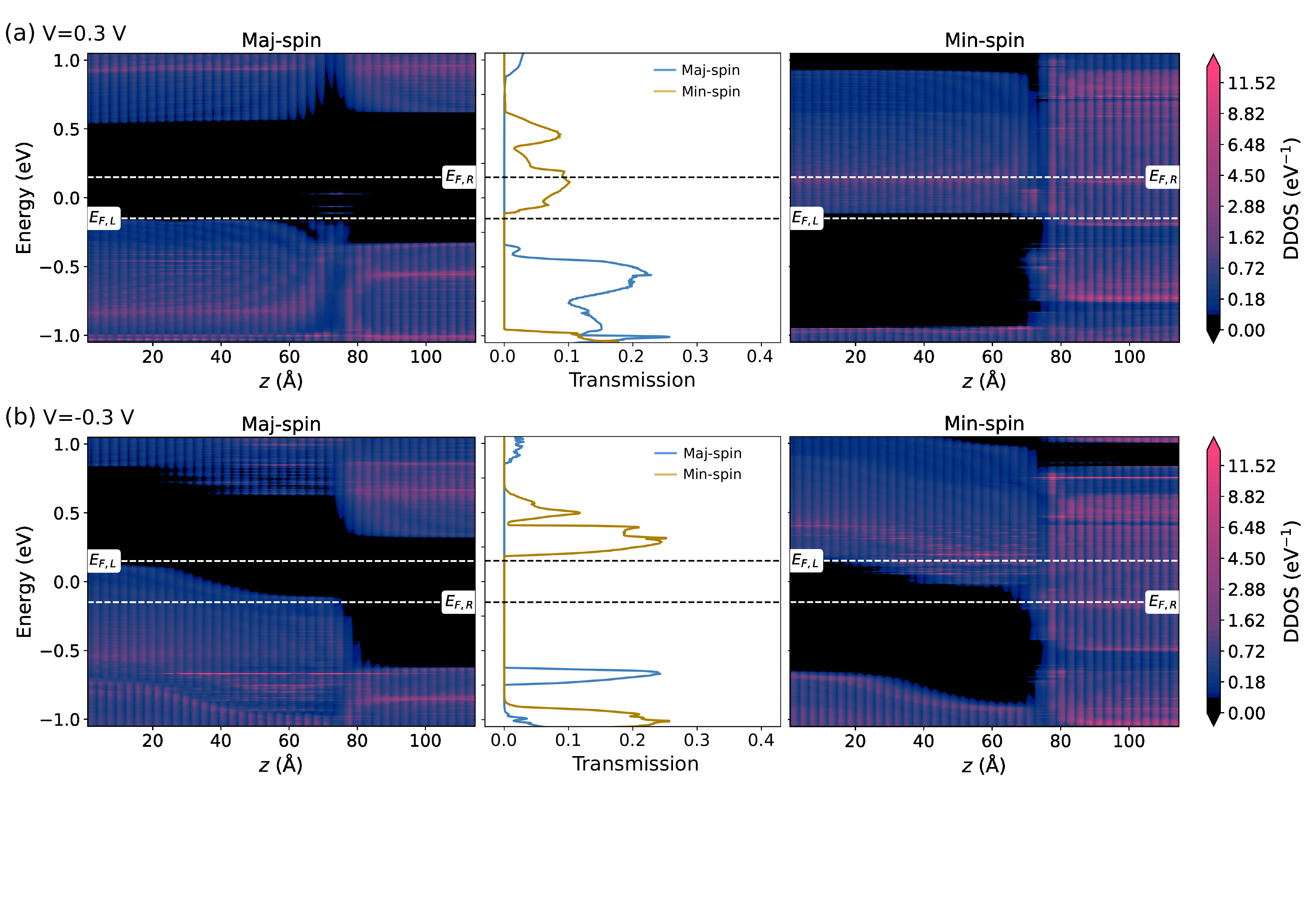}
\end{center}
\vspace*{-2.0cm} 
\caption{ (a) The projected device density of states (DDOS) for the majority (left-hand panel) 
and minority (right-hand panel) electrons for the VS$_2$-Fe/MoS$_2$ Ohmic spin diode (see 
Fig.~\ref{fig6} for the atomic structure of the Ohmic spin diode) under a positive voltage 
of $V=0.3$~V. The middle panel shows the calculated transmission spectrum for the majority 
and minority electrons for the same applied voltage of $V=0.3$~V.  The upper and lower white 
(black) dashed lines indicate the Fermi level of the SGS and HMM electrodes. (b) The same 
as (a) for a negative voltage of $V=-0.3$~V.} 
\label{fig8}
\end{figure*}

Next, we discuss the electronic and transport properties of the OSD under a finite bias voltage. As 
mentioned above, although at zero bias the coupling between the SGS and HMM electrodes is ferromagnetic, 
we find that this coupling changes sign from ferromagnetic to antiferromagnetic under 
a forward bias of about 180~mV. Note that electric field or voltage control of magnetism on the 
nanoscale is highly appealing for the development of nanoelectronic devices with low power 
consumption~\cite{VC_1,VC_2,VC_3}. A voltage-induced interlayer exchange coupling in magnetic tunnel 
junctions has been discussed theoretically via high-voltage tunneling~\cite{theo_1,theo_2,sayed2019electric} and has been
experimentally demonstrated via mobile oxygen vacancies~\cite{newhouse2017voltage}. A sign change in magnetic 
coupling with the bias voltage allows the realization of devices with unique functionalities, 
which will be considered in a separate paper. In Figs.~\hyperref[fig8]{8(a)} and \hyperref[fig8]{8(b)}, 
we show the spin-resolved projected DDOS for bias voltage of 0.3~V (forward bias) and -0.3~V 
(reverse bias), respectively. Note that for the purpose of the demonstration of the OSD concept, we 
constrain the magnetic coupling between electrodes to the ferromagnetic state (parallel orientation) 
in the VS$_2$-Fe/MoS$_2$ junction for bias voltages higher than 180~mV. As one can see, for a forward 
bias, the spin-gapless and the half-metallic behavior is more or less preserved at the interface 
for both spin channels. Just a few new states arise at the interface in the spin-up channel, this being due 
to antiferromagnetic coupling of the single V atom at the interface. The interface V atom possesses a magnetic 
moment of -0.55~$\mu_B$ and variation of the magnetic moments in Fe and other V atoms near the interface 
is also negligible.

On the other hand, for a reverse bias the coupling between electrodes remains ferromagnetic and 
thus the electronic and magnetic structure near the interface is similar to the zero-bias case with the 
exception of the band bending in the VS$_2$ electrode, which take places near the interface region, 
within the 40~{\AA}. Note that the band bending is not linear due to charge transfer as well as the
complex metallic screening of the Fe/MoS$_2$ electrode. Indeed, we observe a flat region of about 
15~{\AA} on the projected DDOS [see Fig.~\hyperref[fig8]{8(b)}] of the junction on the VS$_2$ side 
and then potential drops more or less linearly with distance.

The \textit{I-V} characteristics of the  VS$_2$-Fe/MoS$_2$ OSD can be qualitatively explained on the basis of 
the projected DDOS shown in Fig.~\ref{fig8}. For a forward bias, the OSD is in the on-state, i.e., minority-spin 
electrons from the occupied states of the Fe/MoS$_2$ electrode can flow into the unoccupied states 
of the VS$_2$ electrode due to the Ohmic contact and thus the transmission coefficient takes finite values 
within the bias window, as shown in middle panel of Fig.~\hyperref[fig8]{8(a)}. Meanwhile, for majority-spin 
electrons the transmission coefficient is zero because both Fe/MoS$_2$ and VS$_2$ have no states 
within the voltage window. Thus the on-state current of the OSD is 100\% spin polarized. On the other 
hand, for a reverse bias voltage, no current flows through the OSD since the energy gap in the majority 
spin channel of the Fe/MoS$_2$ electrode prevents the flow of the spin-up electrons from VS$_2$ electrode
into the Fe/MoS$_2$ [see Fig.~\hyperref[fig8]{8(b)} left-hand panel], giving rise to a zero transmission coefficient. 
Similarly, the transmission coefficient for the spin-down channel is also zero due to the energy gap
in the VS$_2$ electrode below the Fermi level.

\begin{figure}[t]
\begin{center}
\includegraphics[scale=0.415]{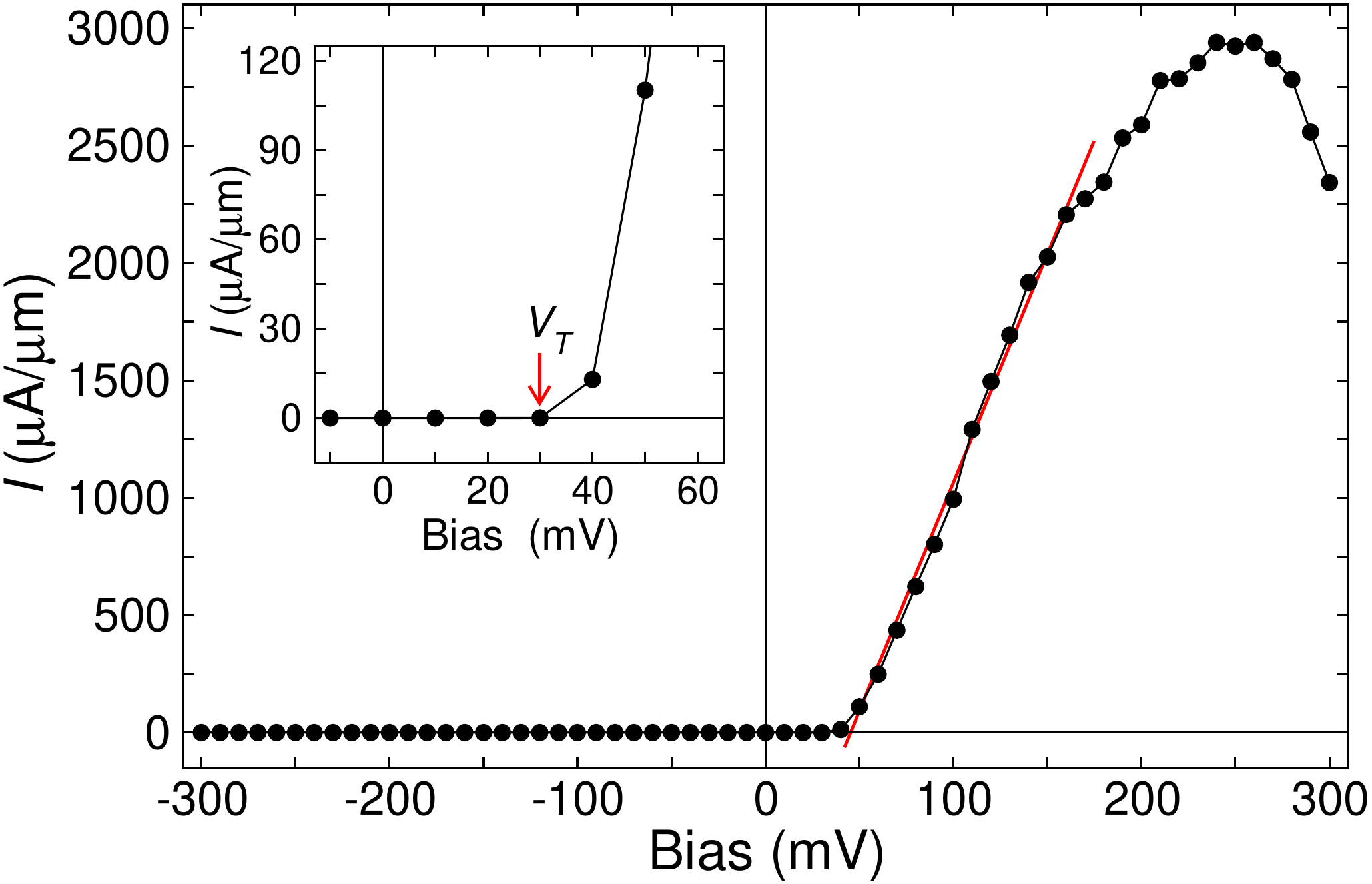}
\end{center}
\vspace{-0.3cm}
\caption{The calculated current-voltage (\textit{I-V}) characteristics of the VS$_2$-Fe/MoS$_2$ 
Ohmic spin diode. The red line shows the Ohmic behavior in the voltage range between 40~mV and 
180~mV. The inset shows an enlargement of the range between -10~mV to 60~mV, to specify the 
threshold voltage $V_T$.}
\label{fig9}
\end{figure}

Fig.~\ref{fig9} shows the calculated \textit{I-V} characteristics of the VS$_2$-Fe/MoS$_2$ OSD at 
zero temperature. In the on-state, the current increases linearly with the applied bias and reaches  
2350~$\mu$A/$\mu$m for a bias voltage of 180~mV, which is the maximum bias voltage at which 
ferromagnetic coupling between electrodes is retained. The VS$_2$-Fe/MoS$_2$ diode possesses 
a very small threshold voltage of about 30~mV, stemming from the indirect band gap of the SGS 
VS$_2$ electrode, which has a conduction-band minimum of 30~meV  [see Fig.~\hyperref[fig6]{6(b)}] 
as discussed before. For bias voltages larger than the 180~mV, the \textit{I-V} curve of the OSD takes a 
plateau shape, i.e., the current first monotonically increases up to 250~mV and then it starts 
to decrease. Such behavior can be attributed to the antiferromagnetic coupling of a single V 
atom at the interface, which now has a gap in the spin-down channel above the Fermi energy, in contrast 
to the rest of the atoms in the junction. This energy gap acts as a small potential barrier, giving 
rise to more reflection of the electrons, and thus it reduces the transmission. Furthermore, in 
the bias-voltage range from 180~mV to 300~mV, the magnetic moment of the V atom at the interface 
increases from -0.2~$\mu_B$ to -0.55~$\mu_B$, which also explains the plateaulike shape of the 
\textit{I-V} curve at higher voltages.

A feature of the proposed VS$_2$--Fe/MoS$_2$ OSD is that it has much higher current-drive 
capability, i.e., in the on state, for a bias voltage of $V=180$~mV the calculated current density 
turns out to be around $I=2350$~$\mu$A/$\mu$m. This value is much higher than the on-state current 
density of devices based on 2D semiconducting materials~\cite{fan2017plane} and lies far above 
the International Roadmap for Devices and Systems (IRDS)~\cite{fan2017plane,IRDS} requirement 
(1350\,$\mu$A/$\mu$m). A negligible turn-on voltage $V_T=30$~mV allows us to detect extremely weak 
signals and thus the OSD might find potential applications in antenna-coupled diode solar 
cells \cite{shanawani2017thz,joshi2016simple,grover2013quantum}. Besides this, the VS$_2$-Fe/MoS$_2$ OSD 
possesses an infinite on:off current ratio at zero temperature. However, as we discussed in 
Sec.~\ref{sec:concept}, thermally excited high-energy electrons would reduce the on:off current 
ratio to finite values. In the Landauer formalism of electronic current, the temperature effects can be taken into account via the Fermi-Dirac distribution function of the left [$f_L (E, V, T)$] and right [$f_R (E, V, T)$] electrodes (see Ref.~\cite{stradi2016general} and thus the on:off current ratio can be caluclated for a given temperature $T$. 
However, this standard treatment should be modified for SGS materials due to their electronic band structure, which is discussed in Sec.~\ref{sec:concept}. Specifically, each spin channel of the SGS material with a proper population of the states above and below the Fermi energy should be taken into account. Since this modification has not been implemented in the current version of the QuantumATK package, we present the \textit{I-V} characteristics only for zero temperature in Fig.~\ref{fig9}.

As mentioned in Sec.~\ref{sec:concept}, the spin degree of freedom brings a certain functionality
to the OSD, which can be dynamically configured by an external magnetic field; thus
the OSD can be used as a switch. However, this is possible only for the OSDs having electrodes 
that couple antiferromagnetically (the AP orientation). So in this case, in reversing the magnetization
direction of one electrode from the antiparallel to the parallel orientation, the OSD switches 
from the on state to the off state like a transistor. This switching can be easily understood 
with the help of the schematic band diagram shown in Fig.~\ref{fig4} and will not be discussed
here. The situation is even more interesting in the case of VS$_2$--Fe/MoS$_2$ OSD, i.e., in 
addition to the voltage-induced switching from the on state to the off state at around $V=180$~mV, 
the OSD can again be switched back to the on state by a weak external magnetic field, making the
VS$_2$--Fe/MoS$_2$ OSD highly appealing for nanoelectronic applications.

Finally, we would like to comment on the effect of low-energy spin excitations on the \textit{I-V} 
characteristics of the VS$_2$-Fe/MoS$_2$ OSD. Apart from the high-energy thermal (non-spin-flip) 
excitations that we discuss in Sec.~\ref{sec:concept}, the temperature affects the electronic and magnetic 
structure of the VS$_2$ and Fe/MoS$_2$ via spin-dependent excitations, i.e., Stoner excitation and 
collective spin waves or magnons. In type-II SGSs such as VS$_2$, electrons can be excited from the 
majority-spin channel to the minority-spin channel via spin flip with almost vanishing energy [see 
Fig.~\hyperref[fig2]{2(c)}]. These excitations are known as single-particle Stoner excitations. 
They can populate the unoccupied minority-spin channel of VS$_2$ just above the Fermi energy. However, 
in VS$_2$, Stoner excitations require a large momentum transfer, since valance-band maximum and conduction-band 
minimum are at different $\mathbf{k}$ points in the Brillouin zone. The former is at the $\Gamma$ 
point, while the latter is close to the $M$ point [for the band structure of VS$_2$, see Fig.~\hyperref[fig6]{6(b)}] 
and thus single-particle spin-flip excitation requires a large momentum transfer and takes place 
at high temperatures close to the Curie temperature $T_C$. Nevertheless, they can populate the unoccupied 
minority channel above the Fermi energy and give rise to the leakage current under a revers bias 
[see Fig.~\hyperref[fig4]{4(c)}]. Moreover, the on:off current ratio will be further reduced. On the other 
hand, in half-metallic Fe/MoS$_2$, Stoner excitations are not allowed due to the existence of the spin 
gap; however, at finite temperatures, electron-magnon interaction can give rise to the appearance of 
nonquasiparticle states within the half-metallic gap just above the Fermi energy \cite{katsnelson2008half}. 
Thus, such states can reduce the spin polarization of the half metallic Fe/MoS$_2$ and affect its transport 
properties. Lastly, in addition to non-spin-flip as well as spin-flip excitations, defects at the interface, 
which destroy the SGS or HMM behavior can also reduce the on:off current ratio of the OSD.

\section{Conclusions}

In conclusion, we proposed a diode concept, which we call the OSD, based on SGS and HMM materials. 
Analogous to the metal-semiconductor junction diode (the Schottky-barrier diode), HMM-SGS junctions act 
as a diode the operation principle of which relies on the spin-dependent transport properties of 
the constituent HMM and SGS materials. We show that the HMM and SGS materials form an Ohmic 
contact under any finite forward bias voltage, giving rise to linear \textit{I-V} characteristics, while 
for a reverse bias the current is blocked due to spin-dependent filtering of the electrons. In 
contrast to the Schottky diode, the proposed OSD does not require doping of the SGS electrode and it 
also does not have a junction barrier; thus it has a zero threshold voltage $V_T$ and an infinite 
on:off current ratio at zero temperature. However, at finite temperatures, non-spin-flip thermally 
excited high-energy electrons as well as low-energy spin-flip excitations can give rise to a leakage 
current and thus reduce the on:off ratio under a reverse bias. As the leakage current is mainly
determined by the size of the band gaps in SGS and HMM electrodes and by a proper choice of large
band-gap materials, the on:off current ratio can be significantly increased in OSDs. Moreover, the spin degree 
of freedom provides a rich configuration space for the \textit{I-V} characteristics of the OSD, which are determined 
by two parameters: (i) the spin character of the gap in HMM and SGS; and (ii) the magnetic coupling between electrodes, 
which allows the dynamical configuration of the diode in the case of antiferromagnetic coupling (antiparallel 
orientation) via an external magnetic field. We show that depending on the magnetic coupling between 
electrodes, the OSD is in the on state either under forward bias (parallel orientation) or under reverse bias 
(antiparallel orientation).

By employing the NEGF method combined with DFT, we demonstrate the rectification characteristics of 
the proposed OSD based on 2D half metallic Fe/MoS$_2$ and spin-gapless semiconducting VS$_2$ planar 
heterojunctions. We find that the VS$_2$–Fe/MoS$_2$ junction diode possesses linear \textit{I-V} characteristics 
for forward bias voltages up to 180~mV and that a bias voltage-induced ferromagnetic to antiferromagnetic
interelectrode coupling then takes place. Such a sign change in magnetic coupling with bias voltage 
allows the realization of devices with unique functionalities, which will be considered in a separate 
paper. Moreover, the VS$_2$–Fe/MoS$_2$ OSD has a much higher current-drive capability ($I=2350$~$\mu$A/$\mu$m)
and a very small threshold voltage of $V_T=30$~mV, which allows us to detect extremely weak signals; thus 
it might find potential applications in antenna-coupled diode solar cells. We expect that our results will 
pave the way for experimentalists to fabricate the OSD based on 2D materials.

\acknowledgements
 E.\c{S} and I.M. acknowledge support from \textit{Sonderforschungsbereich} TRR 227 of the
 Deutsche Forschungsgemeinschaft (DFG) and funding provided by the European Union
 (EFRE), Grant No: ZS/2016/06/79307.
 
E.\c{S} and T.A. contributed equally to this work.

\bibliographystyle{apsrev}
\bibliography{apssamp.bib}

\end{document}